\documentclass[prb,floatfix,twocolumn,showpacs,amsmath,amssymb]{revtex4}
\usepackage{graphicx}
\usepackage{dcolumn}
\usepackage{bm}

\begin{document}

\title{Ground-state properties of the disordered Hubbard model in two dimensions
}
\author{Maria Elisabetta Pezzoli$^{1,2}$ and Federico Becca$^{2,3}$}
\affiliation{
$^{1}$ Department of Physics, Rutgers University, Piscataway, New Jersey 08854,
USA \\
$^{2}$ International School for Advanced Studies (SISSA),
Via Beirut 2, I-34014 Trieste, Italy \\
$^{3}$ CNR-IOM-Democritos National Simulation Centre, Trieste, Italy.
}

\date{\today}
\begin{abstract}
We study the interplay between electron correlation and disorder in the
two-dimensional Hubbard model at half-filling by means of a variational wave 
function that can interpolate between Anderson and Mott insulators. 
We give a detailed description of our improved variational state and explain 
how the physics of the Anderson-Mott transition can be inferred from equal-time
correlations functions, which can be easily computed within the variational
Monte Carlo scheme. The ground-state phase diagram is worked out in both the 
paramagnetic and the magnetic sector. 
Whereas in the former a direct second-order Anderson-Mott transition is 
obtained, when magnetism is allowed variationally, we find evidence for the 
formation of local magnetic moments that order before the Mott transition. 
Although the localization length increases before the Mott transition, we have 
no evidence for the stabilization of a true metallic phase.
The effect of a frustrating next-nearest-neighbor hopping $t^\prime$ is also 
studied in some detail. In particular, we show that $t^\prime$ has two primary 
effects. The first one is the narrowing of the stability region of the 
magnetic Anderson insulator, also leading to a first-order magnetic transition.
The second and most important effect of a frustrating hopping term is the 
development of a ``glassy'' phase at strong couplings, where many paramagnetic 
states, with disordered local moments, may be stabilized.
\end{abstract}


\maketitle

\section{Introduction}\label{sec:intro}

Within independent-electron approaches, all single-particle states are
delocalized and the metallic or insulating behavior is determined by the 
existence of an energy gap between the highest occupied level and the lowest 
unoccupied one. However, whenever the Coulomb interaction becomes dominant
over the kinetic energy, the independent-electron picture fails and electrons 
in the narrow bands close to the Fermi energy become localized. Systems, 
whose insulating character is induced by electron correlations, are called 
{\it Mott insulators}.~\cite{mott} The presence of disorder weakens the 
constructive interference that allows a wave-packet to propagate coherently 
in a periodic potential and may eventually lead to single-particle localization
at the Fermi energy, hence to a further class of insulating materials, 
called {\it Anderson insulators}.~\cite{anderson}
In this context, the conventional one-parameter scaling theory of 
conductance~\cite{gang-of-4} predicts that all solutions of the single-particle
Schr{\oe}dinger equation in a disordered potential are localized
in two dimensions (2D). Therefore, any amount of disorder in 2D drives a 
non-interacting electron system into an Anderson insulator.
The inclusion of electron-electron interaction in weak coupling does not 
modify qualitatively the above conclusion,~\cite{altshuler} although it may 
crucially affect physical properties like the tunneling density of 
states.~\cite{altshuler,altshuler2}
Hence, it is widely accepted that 2D electron systems should always display 
an insulating behavior at sufficiently low temperature or large size no matter 
how weak the disorder is.

Nevertheless, from time to time some indications have appeared that this 
conclusion might not be always correct. Finkel'stein~\cite{finkelstein} and 
Castellani {\it et al.}~\cite{castellani} considered the interplay between 
disorder and interaction by perturbative renormalization group methods 
and showed that, for weak disorder and sufficiently strong interactions, 
a 2D system might scale towards a phase with finite conductivity. However, 
this result was not conclusive since the ``metallic'' region is found to occur 
outside the theory's limits of validity. Indeed, the common feature of these
calculations is the crucial role played by spin fluctuations that grow large 
as the renormalization procedure is iterated. This tendency has been commonly 
interpreted as the emergence of local moments concomitantly with the 
progressive Anderson localization and, in continuum systems, it might signal 
an incipient ferromagnetic instability.~\cite{belitz} In lattice models, 
this is likely to be substituted by a magnetic instability at some wave-vector 
determined by the topology of the Fermi surface.
In this respect, an alternative approach could be to assume from the beginning 
a strong interaction that drives the system into a local moment regime,
e.g., close to a magnetic Mott transition, and then turn on disorder.
However, since it is already difficult to obtain an accurate description of
a clean Mott transition, this approach is very hard to pursue. 
Moreover, assuming the scenario provided by dynamical mean-field theory 
(DMFT),~\cite{georges} one would immediately face with the so called Harris
criterion.~\cite{harris} Indeed, the correlation-length exponent predicted 
within DMFT for a clean Mott transition $\nu \simeq 1/2$ is smaller than 
$2/D=1$ in $D=2$ dimensions.~\cite{rosch,borghi} This fact implies that the 
whole critical behavior has to be profoundly altered by disorder, which 
therefore can not be regarded at all as a weak perturbation.

Only with the groundbreaking experimental work by Kravchenko and
coworkers,~\cite{kravchenko1,kravchenko2} the statement that localization 
always occurs in 2D was really put into question. In fact, Kravchenko has been
the first to observe and claim that, above some critical carrier density, 
high-mobility silicon MOSFETs display a metallic behavior, i.e., a resistivity 
that decreases with decreasing temperature down to the lowest accessible one.
Below this critical density, the behavior of the resistance looks insulating,
thus suggesting that a metal-insulator transition does occur by varying the
density. This experimental finding has renewed the interest in the interplay 
between disorder and interaction and has generated further theoretical 
works along the same direction originally put forward by Finkel'stein, 
in the attempt to clarify some open issues and the applicability of the 
approach.~\cite{castellani2,punnoose} However, the issue of having 
a genuine metallic phases in 2D is controversial and remains under debate, 
from both the experimental and the theoretical point of 
view.~\cite{abrahams,altshuler3,pfeiffer}

An approach where neither the interaction nor the disorder are treated as a
perturbation is therefore required to understand the complex physics of
correlated disordered systems. The simplest model which contains both 
ingredients on the same level is the disordered Hubbard model, which has 
been intensively studied by several numerical methods, such as Hartree-Fock 
calculations in two~\cite{heidarian} and three dimensions,~\cite{imada,imada2}
extended DMFT,~\cite{DMFT,DMFT2,TMT,hofstetter,hofstetter2,kuchinskii}
and quantum Monte Carlo simulations.~\cite{scalettar,ulmke,ulmke2,ulmke3}
However, all these methods have some drawback and only a combined analysis of
complementary techniques can be able to clarify the nature of this challenging
problem. Any approach based on single-particle descriptions, like unrestricted
Hartree-Fock,~\cite{heidarian,bhatt1} can uncover the emergence of 
an insulating gap only by forcing magnetic long-range order. More sophisticated 
approaches, like those based on DMFT,~\cite{DMFT,DMFT2} can in principle 
manage without magnetism,~\cite{TMT,hofstetter,aguiar,aguiar2} but
they usually miss important spatial correlations. The spatial distribution of
local moments and its connection with Griffiths singularities, which may
emerge close to the Mott insulator,~\cite{miranda} have been recently 
discussed within a Brinkmann-Rice approach of the Gutzwiller wave 
function.~\cite{andrade}

In Ref.~\onlinecite{pezzoli}, an improved approach based on a variational
wave function has been proposed to deal simultaneously with the physics of 
Anderson's and Mott's localization. In this work, we present an extensive and 
more complete study of the ground-state properties of the disordered Hubbard 
model in two dimensions at half filling. Moreover, we also discuss in great
details the variational method used in our calculations.
Within our approach it is possible to describe both a direct transition 
between a compressible Anderson insulator and an incompressible paramagnetic 
Mott phase and a transition to a magnetically ordered insulator.
In the first part, we discuss the variational wave function and we explain 
how it is possible to distinguish between Anderson and Mott insulators, by 
means of static correlation functions. In the second part, we describe 
paramagnetic and magnetic transitions. By allowing for magnetic order, we show
the evidence for the formation of local magnetic moments that order 
{\it before} the Mott transition. Finally, we introduce a frustrating 
(next-nearest-neighbor) hopping that favors a glassy behavior with many 
different local minima (and presumably long time scales).

The paper is organized as follows: in Sec.~\ref{sec:model} we present the
method and the variational wave function that is used in our calculations;
in Sec.~\ref{sec:prelim} we show the accuracy of our method;
in Sec.~\ref{sec:noaf}, we show our numerical results for the paramagnetic
sector; in Sec.~\ref{sec:siaf}, we present the results for the magnetic sector;
finally, in Sec.~\ref{sec:conc}, we draw our conclusions.

\section{Model and methods}\label{sec:model}

\subsection{Variational wave function}

We consider the two-dimensional Hubbard model with on-site disorder:
\begin{equation} \label{eq:Ham}
{\cal H} = - \sum_{i,j,\sigma} 
t_{ij} c^{\dagger}_{i,\sigma} c_{j,\sigma} + h.c + \sum_{i} \epsilon_{i} n_{i}
+ U \sum_{i} n_{i,\uparrow} n_{i,\downarrow}
\end{equation}
where $c^{\dagger}_{i,\sigma}$ ($c_{i,\sigma}$) creates (destroys) an 
electron at site $i$ with spin $\sigma$, and $n_i=\sum_\sigma n_{i,\sigma}$ is 
the local density operator. $\epsilon_{i}$ are random on-site energies chosen 
independently at each site $i$ and uniformly distributed between $[-D,D]$, 
$U$ is the repulsive electron-electron interaction and $t_{ij}$ is the
hopping amplitude. In the first part of this work, we consider only a 
nearest-neighbor hopping term $t=1$ and, in the final part, we also add a 
frustrating next-nearest-neighbor term $t^\prime$. We consider 45 degree 
rotated clusters with $N=2 l^2$ sites at half-filling, i.e., with $N_e=N$ 
electrons. 

For $U=0$, the Hamiltonian~(\ref{eq:Ham}) reduces to the Anderson model
for which the ground state is an Anderson insulator for any value of $D$, 
with gapless charge excitation but localized states. On the contrary, 
in the opposite limit $U/t \to \infty$, all charge fluctuations are suppressed, 
the system recovers translational invariance and the ground state becomes a 
Mott insulator. In this work, we study the zero-temperature properties for 
finite disorder $D$ and Coulomb interaction $U$, by using the variational 
Monte Carlo algorithm. Our variational ansatz for the ground state is given by 
\begin{equation}\label{eq:psi}
|\Psi \rangle =  {\cal J} {\cal G} |SD\rangle,
\end{equation}
where $|SD \rangle$ is an uncorrelated Slater determinant that is the ground 
state of a mean-field Hamiltonian:
\begin{equation}\label{eq:Anderson}
{\cal H}_{MF}= - \sum_{i,j,\sigma} {\tilde t}_{ij} 
c^{\dagger}_{i,\sigma} c_{j,\sigma}  + h.c. + \sum_{i,\sigma} 
\tilde{\epsilon}_{i,\sigma} n_{i,\sigma},
\end{equation}
where ${\tilde t}_{ij}=t$ for neighboring sites and 
$\tilde{\epsilon}_{i,\sigma} $ are variational parameters; in addition the
next-nearest-neighbor hopping ${\tilde t}_{ij}={\tilde t}^\prime$ is also used
as a variational parameter in case of a finite value of $t^\prime$ in the
Hamiltonian. We consider both paramagnetic and magnetic properties. 
In the former case, we impose the wave function $|\Psi \rangle$ to be 
paramagnetic, by fixing the variational parameters 
$\tilde{\epsilon}_{i,\downarrow}= \tilde{\epsilon}_{i,\uparrow}$. On the 
contrary, in order to study magnetic properties, we consider a variational wave 
function that may break the spin-rotational symmetry, namely we allow the 
variational parameters to be
$\tilde{\epsilon}_{i,\downarrow} \neq \tilde{\epsilon}_{i,\uparrow}$.

A more general choice of the trial wave function could also contain additional
parameters for all (nearest-neighbor) hopping terms ${\tilde t}_{ij}$ 
(with no translational invariance). Calculations performed on small lattices 
showed that, despite a much larger computational effort, this ansatz leads 
only to slightly lower energies, without modifying the ground-state properties.
Therefore, this possibility will be not considered in the following.

The Gutzwiller factor ${\cal G}$ is defined by
\begin{equation}
{\cal G}=\exp [-\sum_i g_i n_i^2],
\end{equation}
and ${\cal J}$ is a long-range Jastrow term, defined by
\begin{equation}
{\cal J}=\exp \left[-\frac{1}{2} \sum_{i,j}v_{ij} (n_{i}-1) (n_{j}-1) \right].
\end{equation}
While the Gutzwiller factors have been defined with a different parameter 
$g_i$ for each site in order to describe the non-homogeneous character of the 
system, we only consider translational invariant $v_{i,j}=v(|r_i-r_j|)$.
In the following, the Fourier transform of the Jastrow parameters will be
denoted by $v_q$.
All these parameters can be optimized in order to minimize the variational
energy. The choice of taking translational invariant $v_{i,j}$ is done in 
order to reduce the total number of parameters and make the problem tractable 
from a numerical point of view. Moreover, since the Jastrow factor plays a 
primary role in the strong-coupling regime, where the disorder effects are 
suppressed, this choice should not put serious limitations to our variational
wave function.

Summarizing, the  variational parameters are i) the on-site energies 
$\tilde{\epsilon}_{i,\sigma}$ and, for $t^\prime \ne 0$, the hopping 
${\tilde t}^\prime$ of the mean-field Hamiltonian~(\ref{eq:Anderson}), 
ii) the Gutzwiller parameters $g_i$, and iii) the translational invariant
Jastrow parameters $v_{i,j}$. The optimization of the variational state, 
without assuming any particular parametric form, is done by an energy 
minimization, which may deal with a large number (few hundreds) of 
parameters.~\cite{sorella}

\subsection{Correlation functions}

To work out the zero-temperature phase diagram, we make use of the f-sum 
rule that allows us to interpret the small-$q$ behavior of density-density
correlations. Starting from the definition of the average energy of the 
excitations
\begin{equation}\label{eq:fsum}
\Delta_q = \frac{\int d \omega \omega N_q(\omega)}{\int d \omega N_q(\omega)},
\end{equation}
where $N_q(\omega)$ is the dynamical structure factor. By integrating over
frequencies this quantity, we obtain the static structure factor $N_q$
\begin{equation}
N_q = \int \frac{d \omega}{2\pi} N_q(\omega).
\end{equation}
We emphasize that $N_q$ can be easily calculated within the Monte Carlo method
through an equal-time density-density correlation over the ground state
\begin{equation}\label{eq:struct}
N_q = \overline{\langle n_{-q} n_q \rangle} - N_q^{\rm disc},
\end{equation}
where $\langle n_{-q} n_q \rangle = \langle \Psi| n_{-q} n_q |\Psi \rangle$ 
indicates the quantum average and the overbar denotes the disorder average.
Notice that, in order to have a correct definition of the dynamical structure
factor $N_q(\omega)$, we have subtracted the disconnected term
\begin{equation}\label{eq:disc}
N_q^{\rm disc} = \overline{\langle n_{-q} \rangle \langle n_q \rangle},
\end{equation}
which is related to the elastic scattering of electrons and, in a disordered 
system, is finite for generic momenta $q$.~\cite{belitz2}
Notice that, in a disordered system $q$ is no longer a good quantum number but, 
nevertheless, the average over different disorder configurations restores the 
translational invariance. This fact suggests that the density-density structure
factor $N_q$ can be a meaningful quantity to assess physical properties.

After a straightforward calculation, we arrive to a simple expression 
of $\Delta_q$ for small momenta:
\begin{equation}\label{eq:fsum2}
\lim_{q \rightarrow 0} \Delta_q \sim \frac{q^2}{N_q}.
\end{equation}
From this equation, we have the important result that $N_q \sim |q|$ implies 
the existence of gapless charge modes, since $\Delta_q \to 0$ for $|q| \to 0$, 
while if $N_q \sim q^2 $ charge excitations are presumably gapped, since 
$\Delta_q \sim {\rm const}$ for $|q| \to 0$. These criteria have been already 
applied with success to the fermionic and bosonic Hubbard models without 
disorder,~\cite{capello1,capello2} where it has been demonstrated that it is 
possible to describe a true Mott insulator by using a Jastrow-Slater wave 
function. In addition, it has been shown that there is a tight connection 
between $N_q$ and the Jastrow parameters $v_q$. For a clean system, 
a conducting state (metallic or superconducting in the fermionic case
and superfluid in the bosonic one) is characterized by $v_q \sim 1/|q|$ 
in any spatial dimensions, whereas, in order to correctly reproduce the static 
structure factor of a Mott insulator, $v_q$ must be more singular, i.e., 
$v_q \sim 1/q^2$ in one and two dimensions and $v_q \sim 1/|q|^3$ in three 
dimensions.~\cite{capello1}

\begin{figure}
\includegraphics[width=\columnwidth]{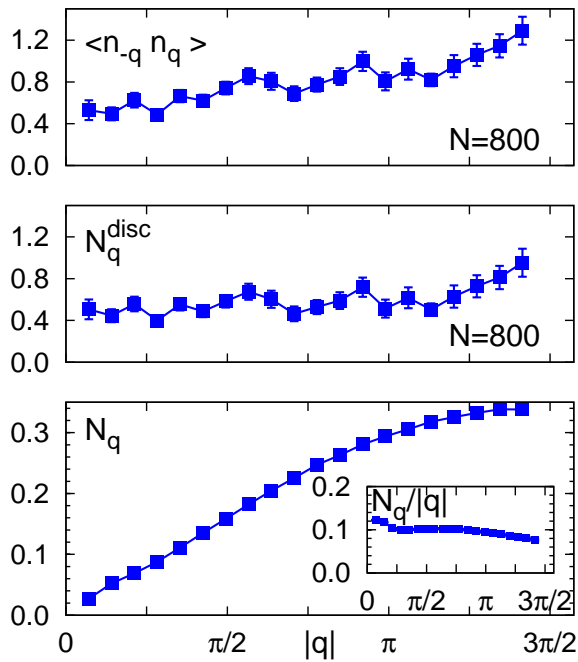}
\caption{\label{fig:struct_U0} 
(Color online) Density-density correlation function 
$\overline{\langle n_{-q} n_q \rangle}$ (upper panel), disconnected term 
$N_q^{\rm disc}$ (middle panel), and 
$N_q=\overline{\langle n_{-q} n_q \rangle} - N_q^{\rm disc}$ (lower panel)
for the non-interacting Anderson insulator. The inset shows $N_q/|q|$, from 
which it is clear that $N_q \sim |q|$. Results are averaged over 48 disorder 
realizations with $D/t=5$ for $N=800$ sites.}
\end{figure}

Let us now turn to the disordered model. Before considering the interacting 
case, we would like to discuss the results for the non-interacting case and 
show that $N_q \sim |q|$ is recovered, in agreement with the fact 
that the ground state is compressible. In Fig.~\ref{fig:struct_U0}, we report
the density-density correlations $\overline{\langle n_{-q} n_q \rangle}$ 
and the disconnected term $N_q^{\rm disc}$ calculated for a system with $N=800$
sites and disorder strength $D/t=5$, averaged over 48 disorder realizations.
It is clear that both quantities are finite for $q \to 0$. However, once we
consider the connected part of the density-density correlations $N_q$,
we have that $N_q \sim |q|$ for $q \rightarrow 0$, in agreement with the 
fact that the Anderson insulator is gapless, see Fig.~\ref{fig:struct_U0}. 
We would like to emphasize that, in contrast to the clean case where 
$N_q \sim |q|$ implies a conducting behavior, here it just indicates a 
compressible state with gapless excitations, but not a metallic character, 
because the single-particle states are localized.

\section{Accuracy of the wave function}\label{sec:prelim}

The variational energy landscape of the disordered Hubbard model may be
characterized by the presence of different local minima. In fact, if we start 
from different points in the parameter space, namely from different values of 
$g_i$, $v_{i,j}$ and $\tilde{\epsilon}_{i,\sigma}$, we may converge to 
different solutions. In the following, whenever we consider the paramagnetic
sector, we impose
$\tilde{\epsilon}_{i,\uparrow}=\tilde{\epsilon}_{i,\downarrow}$ along the
whole optimization procedure. On the other hand, when magnetic solutions are 
allowed, these conditions are relaxed along the Monte Carlo simulation 
and the two on-site energies (for up and down spins) are optimized 
independently. In the latter case, the starting configuration can be taken 
to be either paramagnetic (i.e., with 
$\tilde{\epsilon}_{i,\uparrow}=\tilde{\epsilon}_{i,\downarrow}$) or with a
small staggering of the magnetization (i.e., with 
$\tilde{\epsilon}_{i,\sigma}=\tilde{\epsilon}_i+\sigma(-1)^{|x_i+y_i|}\delta$,
where $\delta$ is a small quantity).
We would like to stress that, even by considering a paramagnetic starting 
point, the converged solution will generally have 
$\tilde{\epsilon}_{i,\uparrow} \ne \tilde{\epsilon}_{i,\downarrow}$.

\begin{figure}
\includegraphics[width=\columnwidth]{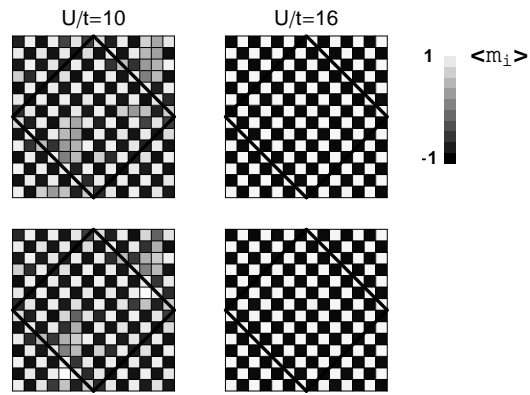}
\caption{\label{fig:tap2s} 
On-site magnetization $\langle m_i \rangle$ for a typical disorder 
configuration with $D/t=5$, $U/t=10$ (left panels) and $U/t=16$ (right panels).
The upper panels correspond to the wave function obtained starting from a 
paramagnetic point, whereas the lower panels correspond to the solution 
obtained from a staggered point. The black contour shows the elementary cell 
of the lattice which is repeated to mimic the infinite lattice with periodic
boundary conditions.}
\end{figure}

\begin{figure}
\includegraphics[width=\columnwidth]{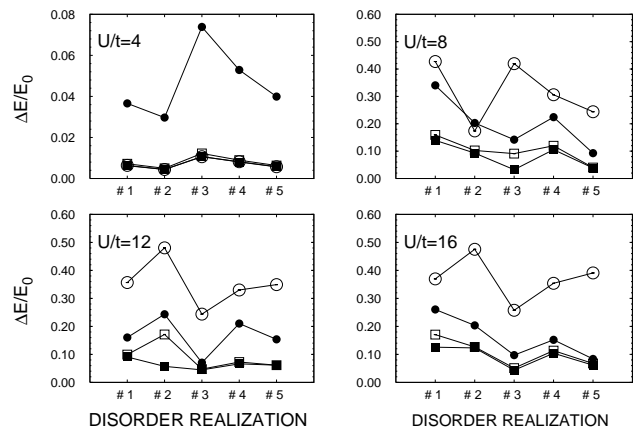}
\caption{\label{fig:accuracy} 
Accuracy of the variational energies $\Delta E = (E_0-E_v)$ (where $E_0$ and 
$E_v$ are the exact and the variational energies, respectively) for different 
wave functions on a $4 \times 4$ lattice with $D/t=5$: the correlated magnetic 
state (full squares), the paramagnetic state (empty circles), the magnetic 
state without Jastrow factors (empty squares), the Hartree-Fock state 
(full circles). The exact ground-state energy is computed by the Lanczos 
algorithm.}
\end{figure}

In contrast to the paramagnetic case, in which the energy landscape usually has
one minimum (i.e., the same parameters are obtained when starting from 
different initializations), when allowing a magnetic wave function different 
local minima may appear. This feature is particularly evident for large enough 
Coulomb repulsion, whereas in the weak-coupling regime we recover a simple 
picture with only one minimum. Remarkably, the appearance of different local 
minima is related to the presence of short-range magnetic correlations. 
In fact, by increasing the on-site Coulomb repulsion, some sites acquire a 
finite magnetization and eventually order, giving rise to the typical staggered
pattern. Whenever local moments are present or the magnetization is very small,
there are different electronic arrangements that give similar energies 
but may be hardly connected by simple single-particle moves, so to define 
metastable local minima. However, especially for the unfrustrated or the 
weakly-frustrated case, the presence of these local minima is not a dramatic 
problem. In fact, generally, all physical quantities, as for instance 
the density-density structure factor $N_q$, give similar results in all these 
cases. Therefore, we can safely conclude that all the optimized states share 
the same physical properties. As an example, Fig.~\ref{fig:tap2s} shows the 
on-site magnetization 
$\langle m_i \rangle = \langle n_{i,\uparrow} - n_{i,\downarrow} \rangle$ 
pattern for the variational wave function optimized both starting from the 
paramagnetic point and from the staggered point. It is evident that there is 
no considerable difference between the two states, although some of the 
magnetization values are slightly different.

We remark that, in most cases, we obtain a lower variational energy by starting
from the paramagnetic point, even if the final converged state is magnetically
ordered. Therefore, although in some particularly delicate cases we considered 
different starting points, we usually initialize the simulation with a 
paramagnetic configuration.

Let us now discuss the accuracy in energy for a $4 \times 4$ lattice, where 
the exact ground state can be calculated exactly by the Lanczos algorithm. 
In particular, we consider four different wave functions: 
i) the magnetic state with on-site Gutzwiller and Jastrow factors 
(that corresponds to our best ansatz), ii) the paramagnetic state with 
Gutzwiller and Jastrow terms, iii) the magnetic state with only Gutzwiller 
projectors, and iv) the magnetic mean-field state  $|SD\rangle$ (i.e., without 
any correlation term).
For $U=0$, the exact ground state wave function can be obtained in all these
cases, implying a very good accuracy also for small but finite values of $U/t$.
For small interactions there is no appreciable differences between paramagnetic 
and magnetic wave functions and for all {\it correlated} states the accuracy 
in the energy, i.e., $(E_0-E_v)/E_0$ (where $E_0$ and $E_v$ are the exact and
the variational energies, respectively), is lower than $1 \%$. 
However, even for $U/t=4$, the Hartree-Fock state, with no Gutzwiller and 
Jastrow factors, give a much worse accuracy than the other three correlated 
wave functions, see Fig.~\ref{fig:accuracy}.
For larger values of the interaction $U$, the situation is different, since the
paramagnetic state is generally worse than the magnetic wave functions,
including the Hartree-Fock state. The accuracy of our best ansatz is about
$10 \%$ (or less) up to very large Coulomb repulsions, which is acceptable
in a disordered model. However, we notice that the long-range Jastrow factor 
is not crucial, and wave functions i) and iii) give comparable energies both 
for small and large interaction values, e.g., $U/t=4$ and $U/t=16$,
see Fig.~\ref{fig:accuracy}. In fact, on the one hand, the on-site Gutzwiller
factor can easily account for the small charge correlations induced by the 
Coulomb repulsion in the weak-coupling regime. On the other hand, for large 
$U/t$, the ground state has strong magnetic correlations and is well described 
with a (gapped) mean-field state. In the intermediate regime, our magnetic 
state with both on-site Gutzwiller projectors and the long-range Jastrow term 
may give a considerable improvement over the other states considered here.
It should be stressed, however, that the Jastrow factor is essential to have a 
{\it paramagnetic} Mott insulator, since in this case the charge gap cannot be 
opened by an uncorrelated Hartree-Fock state. On the contrary, in the magnetic 
case, long-range Jastrow correlations are not strictly necessary to capture 
the correct nature of the ground state, since the charge gap can be naturally 
created by a finite antiferromagnetic mean-field parameter.

\section{Results: the paramagnetic case}\label{sec:noaf}

Let us start our analysis of the disordered Hubbard Hamiltonian~(\ref{eq:Ham}) 
by enforcing a paramagnetic variational wave function, e.g., by fixing the 
constraint $\tilde{\epsilon}_{i,\uparrow} = \tilde{\epsilon}_{i,\downarrow}$ 
for the parameters of the auxiliary mean-field Hamiltonian~(\ref{eq:Anderson}). 
Although this choice is biased, since antiferromagnetic order may be present
for finite electron-electron interaction, it gives significant insights 
into the interplay between disorder and interaction, without any ``spurious''
effect due to magnetism. We apply the f-sum rule to distinguish the 
compressible Anderson insulator from the incompressible Mott insulator, i.e., 
we look at the different behavior of $N_q$ and $v_q$ for different values of 
interaction $U$ and disorder $D$. Here, we take a rather strong disorder 
($D/t=4$, $5$, and $6$) in order to have a localization length that is smaller
than the numerically accessible system sizes.

A detailed analysis of these quantities allows us to identify the Mott 
transition at $U_c^{\rm MI}=(11.5 \pm 0.5) \; t$ for $D/t=5$.~\cite{pezzoli}
Indeed, for small values of the interaction strength, i.e., for 
$U < U_c^{\rm MI}$, we have that $N_q \sim |q|$, whereas $N_q \sim q^2$ in the 
strong-coupling regime $U > U_c^{\rm MI}$. The latter behavior is symptomatic 
of the presence of a charge gap hence of a Mott insulating 
behavior.~\cite{capello1} 
We would like to emphasize that the present results are qualitatively similar 
to those obtained within the clean Hubbard model, with linear coefficient of 
$N_q$ going smoothly to zero at the phase transition, indicating that the 
transition is likely to be continuous. We mention that the Fourier transform 
$v_q$ of the optimized Jastrow parameters $v_{i,j}$ is also compatible with
a Mott transition at the same value of the interaction strength, with its
small-$q$ behavior changing from $v_q \sim 1/|q|$ to $v_q \sim 1/q^2$ across
the Mott transition.~\cite{pezzoli}

\begin{figure}
\includegraphics[width=\columnwidth]{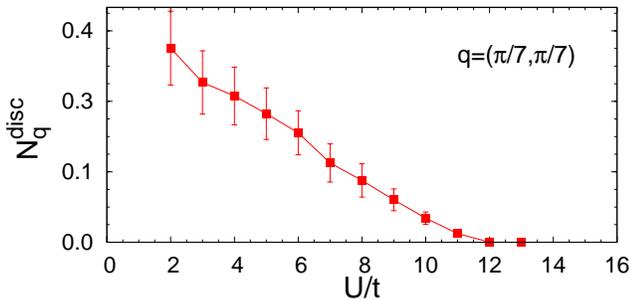}
\caption{\label{fig:orderparameter} 
(Color online) Disconnected part of the density-density correlation function 
$N_q^{\rm disc}$ for the smallest $q$ as a function of the interaction $U$ for 
$D/t=5$ and $N=98$ sites.}
\end{figure}

\begin{figure}
\includegraphics[width=\columnwidth]{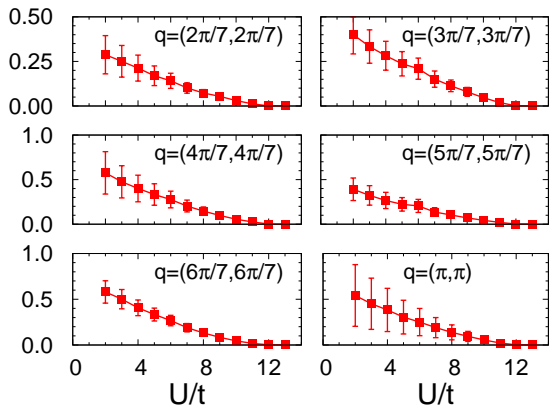}
\caption{\label{fig:tuttiq}
(Color online) The same as in Fig.~\ref{fig:orderparameter} for different 
momenta.}
\end{figure}

Most interestingly we find that $N_q^{\rm disc}$ goes continuously to zero at 
the phase transition, see Fig.~\ref{fig:orderparameter}. This fact allows us 
to identify a simple and variationally accessible quantity to distinguish 
between an Anderson insulator and a Mott insulator. We would like to stress 
that this behavior is not restricted to $q \to 0$, but we recover a similar 
trend also for finite momenta, although disorder fluctuations are larger for 
larger $q$ vectors, see Fig.~\ref{fig:tuttiq}. These results demonstrate that 
charge fluctuations are strongly suppressed and eventually become {\it local} 
as $U/t$ increases.

\begin{figure}
\includegraphics[width=\columnwidth]{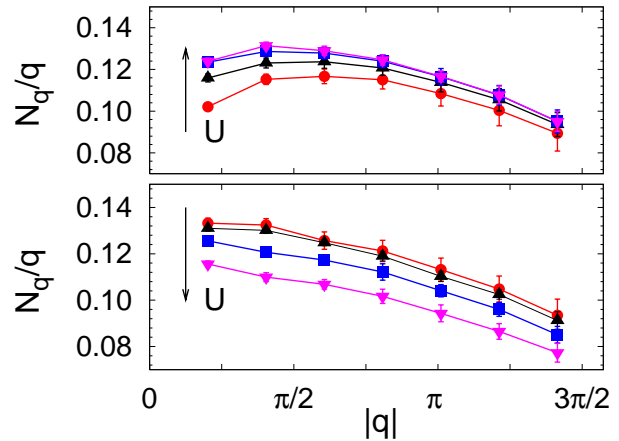}
\caption{\label{fig:nonmono} 
(Color online) Static structure factor $N_q$ divided by $|q|$ as a function 
of $|q|$, for different values of the interaction $U$. 
Upper panel: $U/t=2$ (circles), $3$ (upward triangles), $4$ (squares), 
and $5$ (downward triangles). Lower panel: $U/t=6$ (circles), 
$7$ (upward triangles), $8$ (squares), $9$ (downward triangles). 
Calculations have been done for $D/t=5$ and $N=98$ sites.}
\end{figure}

\begin{figure}
\includegraphics[width=\columnwidth]{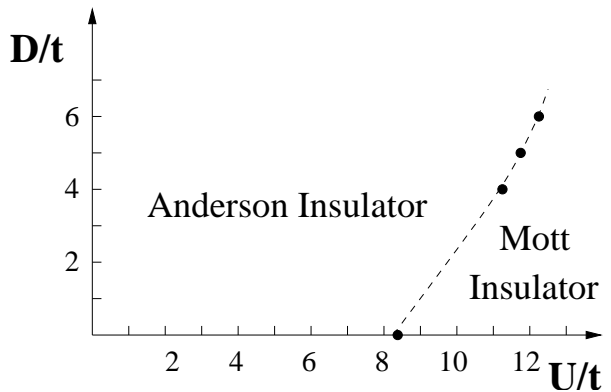}
\caption{\label{fig:phase} 
Phase diagram for the disordered Hubbard model in the paramagnetic sector.}
\end{figure}

All these results suggest that disorder is strongly suppressed in the 
regime of strong correlations. This fact is also corroborated by the
calculation of the variance of the distribution of the on-site energies 
$\tilde{\epsilon}_{i,\sigma}$, which was shown in Ref.~\onlinecite{pezzoli}.
Previous Monte Carlo calculations have shown that a repulsive interaction
may screen the local energies, thereby generating an effectively weaker
random potential.~\cite{scalettar} At the mean-field level~\cite{herbut} and
beyond,~\cite{DMFT2} the screening effects have been widely discussed.
The Hartree-Fock state leads to a disorder screening only for moderate 
interactions, while it gives almost unscreened on-site energies in the 
strongly-correlated regime. On the contrary, our correlated variational 
approach is able to capture the correct physics also for large interaction $U$,
where the disorder, though finite, is highly suppressed. The redistribution 
of on-site energies leads to a decreased localization of the electronic 
state at the Fermi level. Nevertheless, the single-particle eigenstates are
always localized, even though a very large localization length may develop.
Within our variational approach, a important question is whether the action 
of the Gutzwiller correlator and the Jastrow factor could turn a localized 
$|SD \rangle$ into a delocalized many-body state $|\Psi\rangle$. Unfortunately,
the variational method does not give access to dynamical quantities and,
therefore, we cannot make a definite statement. Nevertheless, we tend to 
believe that such a transmutation of a localized $|SD \rangle$ into a 
delocalized $|\Psi\rangle$ is unlikely. In any case, the previous results 
show an increase in the ``metallicity'' of the ground state, with a partial 
screening of disorder.
This result is in agreement with the fact that the linear slope of $N_q$ has 
a non-monotonic behavior as a function of $U$, showing a peak for $U/t \sim 7$ 
that indicates an accumulation of low-energy states around the Fermi energy, 
see Fig.~\ref{fig:nonmono}. In fact, the linear slope of $N_q$ is related
to the compressibility of the system. Nevertheless, it has to be noticed that 
the Slater determinant $|SD \rangle$ is the ground state of a mean-field 
Hamiltonian that always describes non-interacting electrons with on-site 
disorder, no matter how large the Coulomb repulsion is. 
Therefore, even though the single-particle eigenstates may have a very long 
localization length because of the suppression of the effective on site 
disorder, yet this length remains finite in two dimensions. 

The analysis of the density-density correlation function, the Jastrow factor,
and $N_q^{\rm disc}$ for different values of $D/t$ allows us to draw the 
paramagnetic phase diagram in the $(U,D)$ plane, see Fig.~\ref{fig:phase}. 

\begin{figure}
\includegraphics[width=\columnwidth]{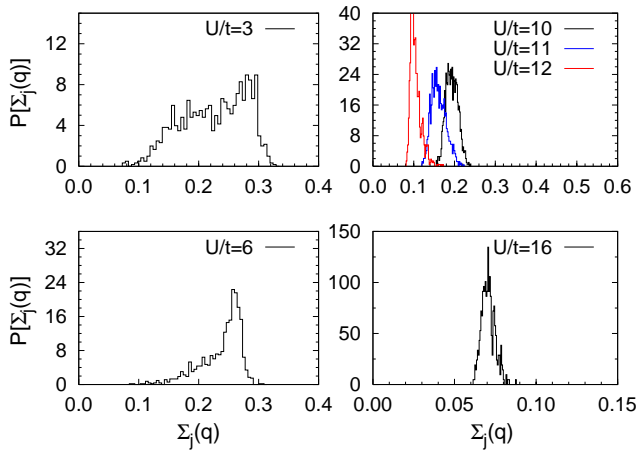}
\caption{\label{fig:local1} 
(Color online) Distribution of $\Sigma_j(q)$ evaluated at $q=(\pi/7,\pi/7)$. 
Calculations are done for $D/t=5$ and $N=98$ sites.}
\end{figure}

\begin{figure}
\includegraphics[width=\columnwidth]{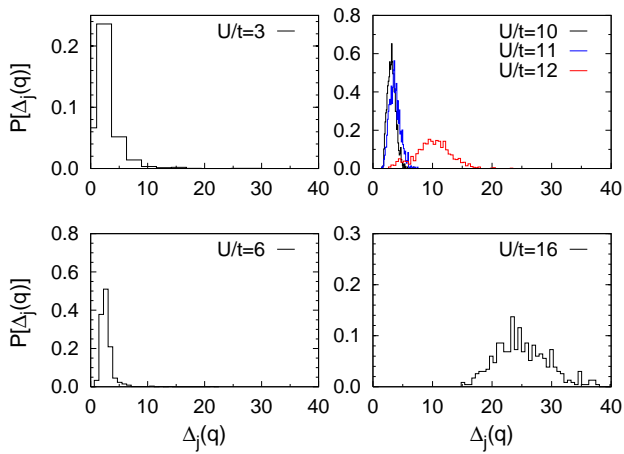}
\caption{\label{fig:local2} 
(Color online) The same as in Fig.~\ref{fig:local1} but for the local gap 
$\Delta_j(q)$.}
\end{figure}

Finally, in order to gain a deeper understanding on the local behavior, namely 
how each single site behaves across the Anderson-Mott transition, we introduce 
a {\it local} f-sum rule:
\begin{equation}\label{localgap}
\Delta_j(q) = \frac{\Sigma_j(q)}{N_j(q)}.
\end{equation} 
Here $N_j(q)$ defines the {\it local} static structure factor:
\begin{equation}\label{localnj}
N_j(q) = \frac{1}{N} \sum_i \langle n_i n_j \rangle_{\textrm{conn}} 
e^{\imath q (R_i-R_j)},
\end{equation}
where $\langle n_i n_j \rangle_{\textrm{conn}}$ represents the quantum average
of $n_i n_j$ after subtracting the disconnected term 
$\langle n_i \rangle \langle n_j \rangle$. In addition, $\Sigma_j(q)$ is 
related to the {\it local} kinetic energy
\begin{equation}
\Sigma_j(q) = -2t \sum_{\langle i \rangle_j,\sigma} 
\langle c^\dag_{i,\sigma} c_{j,\sigma} + h.c. \rangle 
(e^{\imath q (R_i-R_j)} -1)
\end{equation}
where $\langle i \rangle_j$ indicates the sum over the nearest neighbors of
the site $j$. Both $\Sigma_j(q)$ and $N_j(q)$ can be easily evaluated in the
variational Monte Carlo scheme, since they require the computation of 
equal-time correlations.

The limit for small momenta of Eq.~(\ref{localgap}) gives important insights 
into the {\it local} gap, making it possible to understand if at the Mott 
transition all sites become localized simultaneously, or non-homogeneous 
fluctuations are still present. In Fig.~\ref{fig:local1}, we report the 
distribution of $\Sigma_j(q)$, evaluated at the smallest value of the $q$ 
available within a 98-site lattice, i.e., $q=(\pi/7,\pi/7)$.
The average value of $\Sigma_j(q)$ slightly increases by increasing $U$ and it
has a maximum in the regime $U \sim D$ (similarly to what happens to $N_q/|q|$,
see Fig.~\ref{fig:nonmono}). The distribution $P[\Sigma_j(q)]$ is rather large
for small $U/t$ and shrinks when the interaction strength is increased, 
in agreement with the fact that disorder is suppressed by interaction. 
However, even very close to the Mott transition, the variance of 
$P[\Sigma_j(q)]$ stays quite large, indicating that a considerable number of 
sites still has large fluctuations. This fact can be interpreted as a two-fluid
behavior, where a fraction of sites can be regarded as localized particles,
whereas the remaining ones behave like in the Anderson insulator. Therefore,
the Mott transition is driven by only a fraction of the total number of sites.
For $U > U_c^{\rm MI}$, we recover a situation where all sites can be 
ascribed to the Mott phase and $P[\Sigma_j(q)]$ has a very sharp peak, with 
a small variance. The distribution of $N_j(q)$ is very similar to the one of
the local kinetic term. By contrast, the distribution of the local
gap $\Delta_j(q)$ is rather narrow for $U<U_c^{\rm MI}$, where for small 
momenta $\Delta_j(q) \sim 0$, due to a vanishing gap in the Anderson phase,
see Fig.~\ref{fig:local2}. Nevertheless, the distribution has very long 
tails (with very small weight), which are related to disorder fluctuations; 
these tails tend to be suppressed by increasing the interaction $U$. 
For $U > U_c^{\rm MI}$, a charge gap opens up in the average density of states 
but the size of the gap turns out to be different from site to site, which 
implies a rather broad distribution, see Fig.~\ref{fig:local2}.

\section{Results: the magnetic case}\label{sec:siaf}

On the square lattice at half filling, in the absence of frustration and 
disorder, an arbitrarily weak repulsive Hubbard interaction $U$ is able to 
induce long-range antiferromagnetic order. In fact, in this case, the presence
of a perfect nesting in the Fermi surface implies a diverging susceptibility 
at $Q=(\pi,\pi)$ that, in turn, opens a finite gap at the Fermi level. 
Therefore, the ground state is a band insulator for any finite value of the 
interaction $U>0$. By contrast, in the presence of a local random potential, 
the charge gap may be filled by (localized) energy levels, possibly destroying 
the long-range magnetic order. Here, we address the important problem of 
the competition between Anderson localization and magnetic order by using our 
improved variational approach and allowing for a magnetic Slater determinant 
$|SD\rangle$ with 
$\tilde{\epsilon}_{i,\downarrow} \ne \tilde{\epsilon}_{i,\uparrow}$.
First, we discuss the phase diagram of the disordered Hubbard 
model~(\ref{eq:Ham}) with only a nearest-neighbor hopping $t$. In this case, 
a finite value of the interaction $U_c^{\rm AF}$ is needed to have the onset 
of long-range magnetic order: below $U_c^{\rm AF}$ the system is described 
by a standard paramagnetic (compressible) Anderson insulator, above 
$U_c^{\rm AF}$ a finite antiferromagnetic order parameter develops; however,
the excitation spectrum remains gapless and the system is compressible.
By further increasing the interaction $U$, i.e., for $U>U_c^{\rm MI}$ the 
ground state undergoes a second phase transition to an incompressible 
antiferromagnetic insulator with a finite charge gap. Interestingly, in the 
paramagnetic Anderson insulator, local moments with a finite value of 
$\langle m_i \rangle = \langle n_{i,\uparrow}-n_{i,\downarrow} \rangle$ 
develop, suggesting that itinerant electrons may not be able to fully screen 
magnetic impurities created by disorder. 

In the last part, we add a next-nearest-neighbor (frustrating) hopping 
$t^\prime$. Also in this case, we show that the Mott insulating phase is 
always accompanied by magnetic order, although with a sufficiently large ratio 
$t^\prime/t$ many local minima appear in the energy landscape, with competing 
magnetic properties. In particular, we find that the lowest-energy solution 
displays magnetic long-range order but many other disordered states with 
localized moments may be stabilized.

\subsection{Magnetic phase diagram}

Let us now consider large systems. In order to assess the magnetic properties,
we define the total magnetization:
\begin{equation}\label{eq:mstaggered}
M = \frac{1}{N} \sum_i e^{\imath q R_i} \langle m_i \rangle,
\end{equation}
where $m_i = n_{i,\uparrow}-n_{i,\downarrow}$.
In analogy with the clean model, also in presence of disorder, by increasing 
the electron-electron repulsion, there is a tendency toward magnetic order at 
$Q=(\pi,\pi)$, and, therefore, we concentrate on this value of the momentum.

\begin{figure}
\includegraphics[width=\columnwidth]{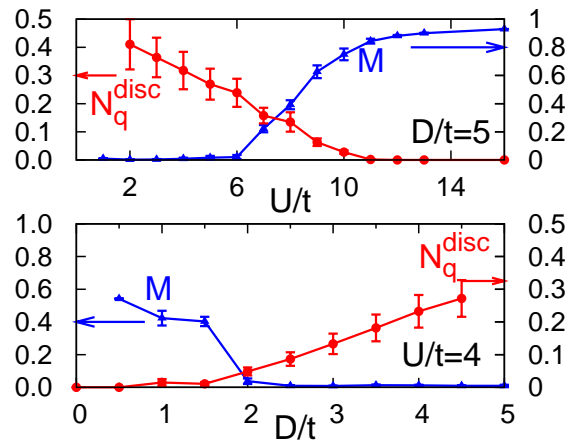}
\caption{\label{fig:magnphase} 
(Color online) Staggered magnetization $M$ and the disconnected term of the 
density-density correlations $N_q^{\rm disc}$ as a function of $U$ for 
disorder $D/t=5$ (upper panel) and as a function of $D$ for $U/t=4$ 
(bottom panel). All calculations have been done for $N=98$ sites.}
\end{figure}

\begin{figure}
\includegraphics[width=\columnwidth]{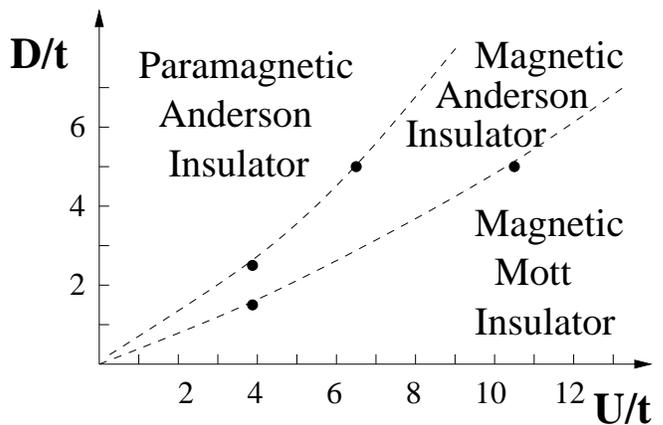}
\caption{\label{fig:phase2} 
Phase diagram for the disordered Hubbard model in the magnetic sector.}
\end{figure}

In Fig.~\ref{fig:magnphase}, we report our results for $D/t=5$ and
different values of the Coulomb repulsion and for $U/t=4$ and various disorder 
strength. Fixing $D/t=5$, we find that $U_c^{\rm AF}=(6.5 \pm 0.5) \; t$,
since the magnetization is finite for $U > U_c^{\rm AF}$ whereas it vanishes
for $U < U_c^{\rm AF}$. Moreover, as we discussed in the previous section, 
the information about charge fluctuations can be worked out from either $N_q$ 
or the disconnected term $N_q^{\rm disc}$. We obtain that 
$N_q^{\rm disc}$ vanishes at $U_c^{\rm MI}=(10.5 \pm 0.5) \; t$, signaling the 
opening of a charge gap, see Fig.~\ref{fig:magnphase}. Remarkably, there 
is a finite region in which both the magnetization and the compressibility 
fluctuations are finite. This fact implies a stable regime that shows 
antiferromagnetic long-range order without a charge gap. We notice that this 
intermediate phase is much reduced when considering $U/t=4$ and vary the 
disorder strength, see Fig.~\ref{fig:magnphase}. In this case, we can estimate
that $D_c^{\rm AF}=(2.5 \pm 0.5) \; t$ and $D_c^{\rm MI}=(1.5 \pm 0.5) \; t$.
These results lead to the phase diagram sketched in Fig.~\ref{fig:phase2}.
For $U=0$ the system is a (paramagnetic) Anderson insulator for every finite 
disorder $D>0$. Instead, for $D=0$ the ground state is a Mott insulator with 
antiferromagnetic order for every $U >0$. When both disorder and interaction 
are finite, there is an intermediate phase between the paramagnetic Anderson
insulator and the antiferromagnetic Mott insulator. This phase is characterized
by long-range magnetic order, but also by a finite compressibility. 
Although some authors identified this phase with a metal,~\cite{heidarian}
we do not find any evidence in favor of it (see below).

\begin{figure}
\includegraphics[width=\columnwidth]{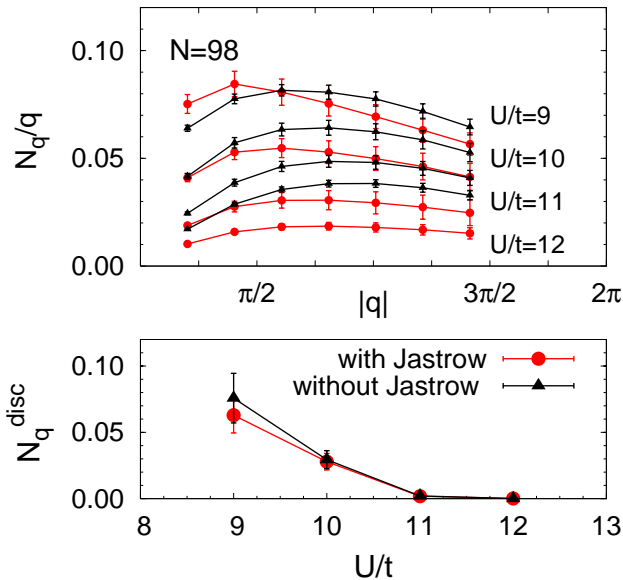}
\caption{\label{fig:afnojas} 
(Color online) Comparison between the results obtained with the full
variational wave function (full circles), containing both Gutzwiller and 
Jastrow terms, and the one obtained with no Jastrow factor but only the 
Gutzwiller projector (full triangles). $N_q$ is shown in the upper panel 
and $N_q^{\rm disc}$ in the lower panel.}
\end{figure}

Let us now consider in more detail the nature of the Anderson-Mott transition 
that emerges from our variational approach, once  we allow for spin-rotational 
symmetry breaking. We would like to remind the reader that, in the paramagnetic 
case, the Mott insulator can be obtained only thanks to a singular Jastrow 
factor, i.e., $v_q \sim 1/q^2$. In this case, the (paramagnetic) mean-field
Hamiltonian~(\ref{eq:Anderson}) is always gapless and the charge gap opens
because of the strong correlations induced by the Jastrow term. In the magnetic
case instead, two different mechanisms can open a gap: i) the long-range 
charge correlations induced by the Jastrow factor and ii) the onset of 
long-range antiferromagnetic order due to a staggering of the
$\tilde{\epsilon}_{i,\sigma}$'s. For $U < U_c^{\rm MI}=(10.5 \pm 0.5) \; t$, 
the static structure factor behaves like $N_q \sim |q|$ and the Fourier 
transform of the Jastrow parameters like $v_q \sim 1/|q|$; on the other hand, 
for $U > U_c^{\rm MI}$, we have that $N_q \sim q^2$ and $v_q \sim 1/q^2$.
Thus in the intermediate phase with long-range magnetic order and finite 
compressibility, $N_q \sim |q|$ and $v_q \sim 1/|q|$.
In order to understand which is the most relevant ingredient that opens the
charge gap, we calculate the structure factor $N_q$ and the disconnected 
term $N_q^{\rm disc}$ close to Mott transition for the full variational
wave function $|\Psi \rangle = {\cal J} {\cal G} |SD \rangle$ and for 
another state that only contains Gutzwiller terms, i.e., 
$|\Psi_g \rangle= {\cal G} |SD \rangle$. The Slater determinant 
$|SD \rangle$ is independently optimized in the two cases. The results for 
$N_q$ and $N_q^{\rm disc}$ are reported in Fig.~\ref{fig:afnojas} and indicate
that the behavior for the two states is very similar; even the critical value 
$U_c^{\rm MI}$ for the Mott transition is the same in the two cases.
Therefore, we can conclude that, in the magnetic case, the charge gap opens 
mainly because of the presence of the mean-field order parameter. 
However, within the correlated wave function $|\Psi \rangle$, the Jastrow 
parameters still behave like $v_q \sim 1/q^2$ in the Mott phase.

\begin{figure}
\includegraphics[width=\columnwidth]{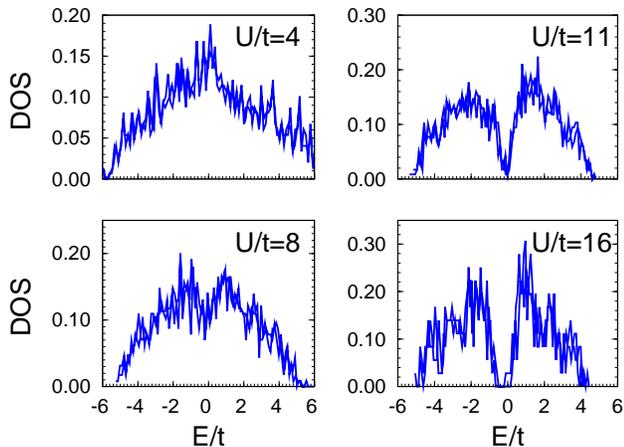}
\caption{\label{fig:meanfDOS} 
(Color online) Evolution of the density of states (DOS) of the auxiliary 
mean-field Hamiltonian as a function of the interaction $U$ for $N=98$ sites. 
The case with spin dependent on-site energies $\tilde{\epsilon}_{i,\sigma}$ 
is considered.}
\end{figure}

In summary, the following scenario emerges: in the intermediate phase with 
antiferromagnetic order but finite compressibility, the mean-field density 
of state is large at the Fermi level, however all states are localized; 
by increasing the interaction $U$, the Jastrow factor becomes stronger and, 
at the same time, there is a suppression of the mean-field density of states 
at the Fermi level; by further increasing $U$, a single-particle gap opens and 
the system becomes incompressible, see Fig.~\ref{fig:meanfDOS}.
We remark that the tendency towards metallicity for intermediate values of 
$U/t$ is suppressed by the presence of magnetic order: this can be seen by 
noticing a reduced density of states at the Fermi level for $U/t \sim 8$.
In fact, for small values of the interaction, the localization length is short 
because of the strong disorder; then it becomes larger for higher values of 
the interaction $U$ due to the disorder screening and then, at the 
antiferromagnetic transition, it decreases again. Moreover, we notice that 
the single-particle wave functions are more localized in the magnetic case 
than in the paramagnetic one, even in the regime of maximum ``delocalization'',
i.e., $U \sim D$. Unfortunately, a full analysis of size scaling of the 
localization length (by considering the inverse participation ratio, as used 
in Ref.~\onlinecite{heidarian}) is very hard, because the available sizes
do not allow us to reach definitive conclusions.

\begin{figure}
\includegraphics[width=\columnwidth]{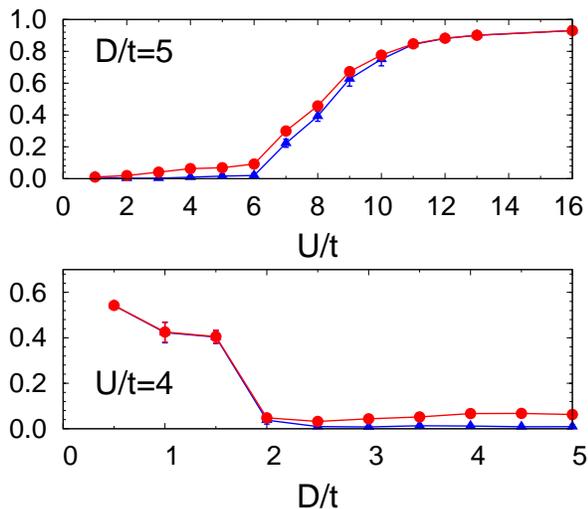}
\caption{\label{fig:2tot} 
(Color online) Staggered magnetization $M$ (blue triangles) and fluctuations 
of the local magnetization $M_L$ (red circles) as a function of $U$ for 
disorder $D/t=5$ (upper panel) and as a function of $D$ for $U/t=4$ 
(bottom panel). Calculations have been done for $N=98$ sites.}
\end{figure}

\begin{figure}
\includegraphics[width=\columnwidth]{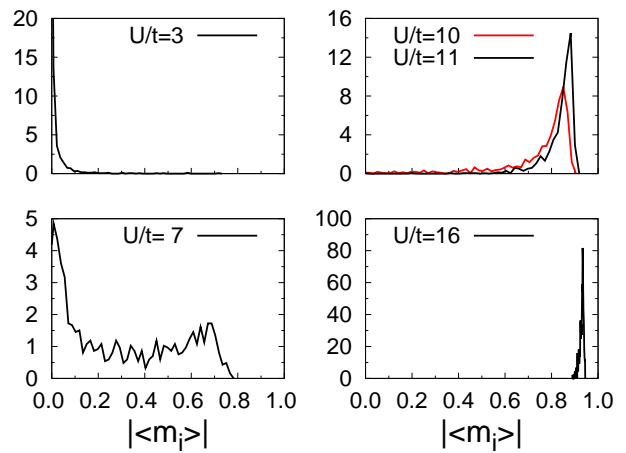}
\caption{\label{fig:distrmom} 
(Color online) Probability distribution of the of the absolute value of the
on-site magnetization $|\langle m_i \rangle|$ for $N=98$, $D/t=5$ and 
different values of the interaction $U/t$.}
\end{figure}

Let us now focus on the development of local magnetic moments in presence
of the Coulomb interaction, in order to demonstrate that they may appear
before the metal-insulator transition. For this issue, we consider the 
following quantity, which is related to the fluctuations of the local 
magnetization
\begin{equation}\label{eq:mlocal}
M_L = \sqrt{ \frac{1}{N} \sum_i \langle m_i \rangle^2}.
\end{equation}
In a paramagnetic state with local moments, namely a state in which some sites 
have an on-site magnetization $\langle m_i \rangle \neq 0 $, the total 
staggered magnetization is vanishing, i.e., $M=0$, while $M_L$ is finite. 
On the contrary, in the antiferromagnetic phase $M_L \simeq M$. 
Therefore, by comparing $M_L$ and $M$, it is possible to have a good feeling 
on the presence of local moments in the ground state. 
In Fig.~\ref{fig:2tot}, we show the staggered magnetization $M$ and 
$M_L$ both for $D/t=5$ and different values of the interaction $U$ and for 
$U/t=4$ and different disorder strengths $D$. We find that for 
$U > U_c^{\rm AF}= (6.5 \pm 0.5) \; t$ [and for 
$D < D_c^{\rm AF}=(2.5 \pm 0.5) \; t$] the two magnetization values are very
close, while in the paramagnetic phase we observe that $M_L > M \simeq 0$.
This fact suggests a magnetically disordered phase in which the on-site
magnetization $\langle m_i \rangle$ is finite for some sites.
We identify those sites with $\langle m_i \rangle \neq 0$ as local magnetic
moments. The existence of such moments can be extracted from the pattern of 
the on-site magnetization $\langle m_i \rangle$ (as shown in 
Ref.~\onlinecite{pezzoli}) or from the probability distribution of
$|\langle m_i \rangle|$, see Fig.~\ref{fig:distrmom}.
Here, we observe that for small interaction values, $U/t \sim 3$, the
distribution has a narrow peak in correspondence of $|\langle m_i \rangle|=0$,
however, at the same time, it has long tails indicating the presence of local
moments. In this regime, the ground state is an Anderson insulator with a large
number of paramagnetic sites and $\langle n_i \rangle=0,1,2$.
For $U \simeq U_c^{\rm AF}$ the distribution is spread between 0 and
0.8, with a peak in correspondence of $|\langle m_i \rangle| \sim 0$.
This fact highlights the coexistence of paramagnetic sites with local magnetic 
moments; these sites are not spatially correlated hence long-range magnetism 
is absent. By increasing the interaction strength, the peak at
$|\langle m_i \rangle| \sim 0$ disappears and the one at
$|\langle m_i \rangle| \sim 1$ becomes more pronounced; in this case, local
moments eventually display the typical staggered pattern of N\'eel order.
Nevertheless, charge excitations are still gapless and $N_q \sim |q|$.
Finally, in the Mott insulating phase the distribution has a narrow peak at
$|\langle m_i \rangle|=1$.

\subsection{Frustrating case}

In the previous section, we showed that a disordered system of electrons on
a square lattice undergoes a magnetic transition before becoming a Mott
insulator. Therefore, the Mott insulator is generically accompanied by magnetic
order. However, in disordered materials, one could expect that long-range
order may be strongly suppressed, leading to a {\it bona-fide} Mott transition, 
where the incompressible phase has no magnetic order. In this sense, the 
presence of a next-nearest-neighbor hopping $t^\prime$ may help to approach 
the Mott phase without any ``spurious'' magnetic effects. Indeed, this kind of 
frustrated hopping generates, in the strong-coupling regime, a super-exchange 
term $J^\prime$ that competes with the nearest-neighbor one $J$. 

First of all, we notice that a finite frustrating ratio $t^\prime/t$ 
generates very complicated energy landscapes, with many local minima.
Furthermore, in contrast to the unfrustrated model, where different local 
minima share very similar physical properties, here different starting points 
in the parameter space may give rise to rather different wave functions, 
especially in the intermediate and strong-coupling regimes. Concerning the 
starting point for the energy optimization, we will consider i) a paramagnetic 
point, with $\tilde{\epsilon}_{i,\uparrow}=\tilde{\epsilon}_{i,\downarrow}$,
ii) a staggered point, with $\tilde{\epsilon}_{i,\sigma}=
\tilde{\epsilon}_i + \sigma (-1)^{|x_i+y_i|} \delta$, and iii) a collinear
point, with $\tilde{\epsilon}_{i,\sigma}=\sigma (-1)^{|x_i|}$.
Notice that, in the latter case, the starting choice breaks both 
translational and rotational invariance, so to favor collinear magnetic order 
with $Q=(\pi,0)$, which is suitable for large $t^\prime/t$. Similarly, we can
also consider $\tilde{\epsilon}_{i,\sigma}=\sigma (-1)^{|y_i|}$, which gives
rise to a collinear order with $Q=(0,\pi)$.
In all these cases, each local energy $\tilde{\epsilon}_{i,\sigma}$ (for up
and down spins) is independently optimized, in order to achieve a full energy 
minimization.

\begin{figure}
\includegraphics[width=\columnwidth]{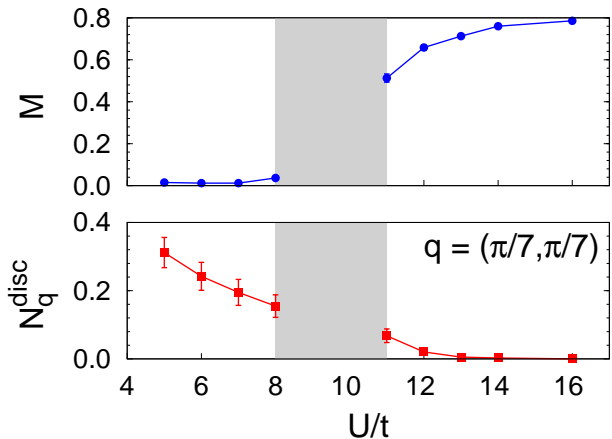}
\caption{\label{fig:collord} 
(Color online) Collinear magnetization $M$ (upper panel) and the 
disconnected term of the density-density correlations $N_q^{\rm disc}$ 
(lower panel) of the lowest energy solution as a function of $U$ for $D/t=5$ 
and $t^\prime/t=1$. The shaded area indicates the region where paramagnetic 
and magnetic solutions have similar energies.}
\end{figure}

For weak and intermediate frustration, the outcome is rather similar to the
case with $t^\prime=0$, except for $U \sim U_c^{\rm AF}$. For instance, 
for $t^\prime/t=0.6$ and $D/t=5$, the paramagnetic Anderson insulator is 
stable for $U < U_c^{\rm AF} \simeq 6 t$. Up to this value of the Coulomb 
interaction, the energy of the converged state does not depend upon the choice 
of the starting point (as for the unfrustrated case) and there is no evidence
for any non-trivial magnetic pattern. On the contrary, for 
$6 < U/t < 12$, different results for the staggered magnetization
are found when we initialize according to i) or ii). In particular, the 
magnetization is strongly suppressed when considering a paramagnetic initial 
condition. These different solutions are very close in energy up to 
$U \simeq 8 t$, even though, usually, the lowest minimum shows long-range order 
and finite compressibility; for larger values of the interactions, i.e., 
$8 < U/t < 12$, the magnetic state gives definitely the best 
energy. By further increasing the Coulomb repulsion, namely for 
$U > U_c^{\rm MI} \simeq 12 t$, the system becomes a Mott insulator with N\'eel 
order at $Q=(\pi,\pi)$. These results suggest that, for a finite frustrating 
ratio, the magnetic transition may become weakly first order (with a
regime in which there is a coexistence of different phases).

The existence of a finite region with nearly degenerate states with different
magnetic properties is even more pronounced in the strong frustration regime.
For $t^\prime=t$, the solution obtained starting from ii) gives always an 
higher energy with respect to i) and iii), demonstrating that the N\'eel state
is clearly disadvantaged. Again, in the weakly-correlated regime,
i.e., $U < U_c^{\rm AF} \simeq 8 \; t$, paramagnetic and magnetic 
starting points give similar energies and magnetization patterns. 
For $U > U_c^{\rm MI} \simeq 13 \; t$, the ground state is a gapped 
insulator with $N_q^{\rm disc}=0$, see Fig.~\ref{fig:collord}. 
Instead, in the intermediate region, for $U_c^{\rm AF} < U < U_c^{\rm MI}$, 
the energy landscape shows different minima, which are nearly degenerate up 
to $U \simeq 11 t$, whereas, for larger interactions, the wave function with 
collinear order gives the best approximation for the ground state, with a 
clear long-range magnetic order, see Fig.~\ref{fig:collord}.

The remarkable feature is that, in a wide range of Coulomb repulsions,
it is possible to find low-energy states just by starting from a paramagnetic
wave function. This choice gives rise to patterns in which most of the
sites have a net magnetization but an overall vanishing magnetic order.
For $U/t \sim 16$, these solutions are incompressible, i.e., 
$N_q^{\rm disc} \sim 0$ and, therefore, may be viewed as disordered Mott 
insulators. By decreasing the interaction strength, these states turn 
compressible, still having a large number of local moments. 
The presence of these metastable solutions, which are almost degenerate
with the magnetically ordered wave function, suggests a sort of 
``spin-glass'' behavior. The existence of a large number of such disordered 
states prevents one to have a smooth convergence to the lowest-energy solution, 
by starting from a generic configuration.

In summary, the frustrating hopping $t^\prime$ has two primary effects. 
The first one is the narrowing of the stability region of the magnetic Anderson
insulator. In addition, we have evidence that the magnetic transition turns 
to be first order, in contrast to the unfrustrated case. The second and most 
important effect of a frustrating hopping term is the development of a 
``glassy'' phase at strong couplings, where many paramagnetic states, with 
disordered local moments, may be stabilized. Although we do not find any 
evidence in favor of the stabilization of a true (non-magnetic) Mott phase, 
this possibility cannot be ruled out in the whole phase diagram.

\section{Conclusions}\label{sec:conc}

In this paper, we have studied, by means of a variational Monte Carlo 
technique, zero-temperature properties of the disordered two-dimensional 
Hubbard model at half filling. 
First of all, we showed that a variational wave function is able to 
describe the Anderson-Mott transition without any symmetry breaking, i.e., 
a transition from a paramagnetic Anderson insulator to a paramagnetic Mott 
insulator. This is achieved thanks to long-range charge correlations 
induced in the wave function by a Jastrow factor. We showed that the 
transition can be easily detected within variational Monte Carlo by looking 
at the behavior of the static structure factor and of the Fourier transform 
of the Jastrow parameters, namely following the same criteria for the
Mott transition in a clean system. Moreover, we found that the disconnected 
term of the density-density correlation function, i.e., 
$\lim_{q \to 0} \overline{\langle n_{-q} \rangle \langle n_q \rangle}$, 
acts as an easily accessible order parameter for the Anderson-Mott transition. 
We found that electron-electron repulsion partially screens disorder: 
for strong interaction electrons feel an effective weak disorder potential 
that should imply an interaction-increased localization length. However, 
once the interaction exceeds a critical value, a gap opens and the model turns 
into a Mott insulator. From our numerical calculations, the ground state is
always insulating, yet, upon increasing the strength of interaction, the 
localization length may have a non-monotonous behavior when we consider the 
full variational wave function.

When magnetism is allowed, a compressible and magnetic Anderson insulating
phase appears between the compressible paramagnetic Anderson insulator
and the incompressible magnetic Mott insulator. When magnetism is not 
frustrated, all transitions are likely to be continuous. On the contrary,
when frustration is included by means of next-nearest-neighbor hopping, 
the paramagnetic to magnetic transition turns first order. 
Moreover, in the magnetic region, it is also possible to stabilize many
paramagnetic solutions with very low energy, suggesting a glassy behavior at 
finite temperature. Indeed, all these paramagnetic states have local moments, 
i.e., magnetic sites that would contribute with a finite $-k_B \ln 2$ term to 
the entropy at finite temperature. The fact that, in this simple 
two-dimensional model, we find local moments in the paramagnetic phase may 
suggest that this is a general feature of disordered systems close to a Mott 
transition.

\acknowledgments
We are particularly indebted to M. Fabrizio for enlightening discussions and
for his careful reading of the manuscript.

\end{document}